\documentclass[12pt]{article}
\newcommand{\dd}{\mbox{\rm d}}
\newcommand{\wg}{\wedge}
\newcommand{\gam}{\gamma}
\newcommand{\dg}{\dagger}

\newcommand{\nnn}{\noindent}
\newcommand{\oo}{\over}
\newcommand{\p}{\partial}

\newcommand{\be}{\begin{equation}}
\newcommand{\bear}{\begin{eqnarray}}
\newcommand{\ear}{\end{eqnarray}}
\newcommand{\ee}{\end{equation}}
\newcommand{\lbl}{\label}
\newcommand{\bi}{\bibitem}
\newcommand{\ci}{\cite}

\newcommand{\vs}{\vspace}

\begin{document}

\

\baselineskip .7cm 

\vs{27mm}

\begin{center}

{\LARGE \bf Clifford Space as a Generalization of Spacetime:
Prospects for QFT of Point Particles and Strings}

\vs{3mm}

Matej Pav\v si\v c

Jo\v zef Stefan Institute, Jamova 39,
1000 Ljubljana, Slovenia

e-mail: matej.pavsic@ijs.si

\vs{6mm}

{\bf Abstract}

\end{center}

\vs{2mm}

The idea that spacetime has to be replaced by Clifford space ($C$-space)
is explored. Quantum field theory (QFT) and string theory are
generalized to $C$-space. It is shown how one can solve the
cosmological constant problem and formulate string theory without
central terms in the Virasoro algebra by exploiting the
peculiar pseudo-Euclidean signature of $C$-space and the Jackiw
definition of the vacuum state. As an introduction into the subject,
a toy model of the harmonic oscillator in pseudo-Euclidean space is studied.

\vs{8mm}

\section{Introduction}

Quantum field theory is a very successful theory, but also enigmatic.
There occur infinities which require renormalization. This
suggests that the theory is not yet complete\footnote{P.A.M. Dirac
expressed such a view, e.g., in a talk presented at the Erice 1982 conference
\ci{Dirac}.}.
An approach whose roots go back to Feynman\footnote{See a nice paper by
Schweber \ci{Schweber}.} \ci{Feynman} employs an invariant evolution parameter
$\tau$, introduced by Fock and Stueckelberg \ci{Stueckelberg}, and
subsequently pursued
by many authors \ci{Horwitz,PavsicBook}.
Another direction of research starts from the idea that spacetime has
to be replaced by a more general space, namely the Clifford space
(shortly, $C$-space) \ci{PavsicBook}--\ci{CliffConform}. It has been found
\ci{Pavsic,PavsicBook} that the Stueckelberg theory with the Lorentz invariant
evolution parameter $\tau$ naturally occurs as being embedded in the
theory based on Clifford space. In this paper we would like to
point out another important aspect of Clifford space. We will show that
quantum field theory generalized to Clifford space
provides a natural way of resolving the notorious
cosmological constant problem. We exploit the property of Clifford space
with signature $(+++ ... --- ...)$, where the number of plus and
minus signs is the same, provided that the underlying spacetime has
Minkowski signature. We find that by using the Jackiw definition
of vacuum \ci{Jackiw}, the concept of $C$-space enables a formulation
of QFT in which zero point energies belonging to positive and negative
signature degrees of freedom cancel out, while preserving the Casimir
effect.

Introduction of $C$-space has consequences for string theory which can be
formulated without central terms in Virasoro algebra even when the
dimension of the underlying spacetime is four. We do not need
a higher dimensional target spacetime for a consistent
formulation of (quantized) string theory. Instead of a higher dimensional
space we have the 16-dimensional Clifford space which also provides
a natural framework for description of superstings and supersymmetry,
since spinors are just the elements of left or right minimal ideals
of Clifford algebra \ci{Teitler}--\ci{Lounesto}, \ci{PavsicBook,Mankoc}.

After a brief review of the concept of Clifford space  we first discuss
a toy model of the harmonic oscillator in pseudo-Euclidean space. We
employ the obvious fact that if a Lagrangian is multiplied by $-1$,
whilst the definitions of momentum and energy are kept the same, namely,
$p_\mu = \p L/{\dot x}^\mu$, $E=p_\mu {\dot x}^\mu - L$, then energy becomes
negative. In such case the criterion for stability is reversed:
the system is stable when its energy has maximum. This is precisely what
happens in a pseudo-Euclidean space: The Lagrangian is a quadratic form
which has the terms with positive and negative signs. Therefore the
expression for energy is also composed of the terms with positive and
negative signs.

We then describe a system of $n$ scalar fields forming a space $M_{r,-s}$,
$n=r+s$, and show that when $r=s$, the zero point energy vanishes. The vacuum
contribution to the stress-energy tensor is zero, and there is no
cosmological constant problem \ci{cosmological1}
in this model\footnote{An eventual small
non-vanishing cosmological constant, as confirmed by recent
observations \ci{cosmological2}, can be a residual
effect of something else, or due to our incomplete understanding of
the dynamical laws at cosmological scales.}. Then we discuss a model
in which the space of
$n$-fields is generalized to the corresponding Clifford space.

Finally, we discuss the string theory and generalize it to $C$-space.

\section{Clifford Space as the Arena for Physics}

Clifford algebra is a very useful tool for description of geometry. Since
Hestenes's seminal books \ci{Hestenes}, there is a growing interest
in using Clifford algebra in physics (see e.g. refs. \ci{Lounesto}--\ci{Moya}),
and generalizing physics from spacetime to Clifford space
\ci{PavsicBook}--\ci{CliffConform}.
An $n$-dimensional flat space (e.g, spacetime) $M_n$ can be described
by means of a complete set of basis vectors $\gam_\mu$, $\mu=0,1,2,...,n-1$,
which satisfy the Clifford algebra relations
\be
     \gamma_\mu \cdot \gamma_\nu \equiv {1\oo 2} (\gamma_\mu
    \gamma_\nu + \gamma_\nu \gamma_\mu) = g_{\mu \nu}
\lbl{2.1}
\ee
This is just the symmetric part of the Clifford (or geometric) product
$\gam_\mu \gam_\nu$. It is equal to the metric $g_{\mu \nu}$. The antisymmetric
part defines a bivector which represents an oriented unit area:
\be
    \gam_\mu \wedge \gam_\nu = {1\oo 2} (\gam_\mu \gam_\nu - \gam_\nu \gam_\mu)
\equiv {1\oo 2} [\gam_\mu, \gam_\nu]
\lbl{2.2}
\ee
This can be continued to the antisymmetrized product of $3,4,..,n$ basis vectors
\be
    \gam_{\mu_1}\wedge \gam_{\mu_2} \wedge  \gam_{\mu_3} =
   {1\oo 3!} [\gam_{\mu_1},\gam_{\mu_2},\gam_{\mu_3}]
\nonumber
\ee
\be   
   \vdots
\nonumber
\ee
\be
     \gam_{\mu_1}\wedge \gam_{\mu_2} \wedge ... \wedge \gam_{\mu_n}
 = {1\oo r!} [\gam_{\mu_1},\gam_{\mu_2},...,\gam_{\mu_n}]
\lbl{2.3}
\ee
An object $\gam_{\mu_1}\wedge \gam_{\mu_2} \wedge ... \wedge \gam_{\mu_r}$
of degree $r$, $1 \le r \le n$, is called a basis {\it multivector}
or $r$-vector. it represents an oriented $r$-dimensional unit area.

A point $P$ in $M_n$ can be associated with a {\it
vector} $x$ joining the coordinate origin $O$ and $P$:
\be
    x = x^\mu \gamma_\mu
\lbl{2.4}
\ee

A generic object is a Clifford number, called also a {\it polyvector}
or {\it Clifford aggregate}, which is a superposition of multivectors\footnote{
Although Hestenes and others use the term `multivector' for a generic
Clifford number, we prefer to call it `polyvector', and reserve the name
`multivector' for objects of definite grade. So our nomenclature
is in agreement with the one used in the theory of differential forms,
where 'multivectors' mean objects of definite grade, and not a superposition
of objects with different grade.}:
\be
    X = \sigma {\underline 1} + x^\mu \gam_\mu + {1\oo 2} x^{\mu_1 \mu_2}
   \gam_{\mu_1 \mu_1} + ... + {1\oo n!} x^{\mu_1 ... \mu_n}
   \gam{\mu_1 ... \mu_n} \equiv x^M \gam_M
\lbl{2.5}
\ee
Here $\gam_{\mu_1 ... \mu_r} \equiv \gam_{\mu_1} \wedge \gam_{\mu_2}
\wedge ... \wedge \gam_{\mu_r}$ and 
\bear
    &&x^M = (\sigma, x^\mu, x^{\mu_1 \mu_2},..., 
  x^{\mu_1 ... \mu_r} )\nonumber \\
   &&\gam_M = ({\underline 1}, \gam_\mu , \gam_{\mu_1 \mu_2},...,
\gam_{\mu_1 ... \mu_r}) \; , \quad \mu_1 < \mu_2 < ... < \mu_r 
\lbl{2.6}
\ear
are respectively coordinates and basis elements of Clifford algebra.

The coordinates $X^{\mu_1 ... \mu_r}$ determine an oriented
$r$-area. They say nothing about the precise form of the $(r-1)$-loop
enclosing the $r$-area. The coordinates 
$\sigma, ~x^{\mu},~ x^{\mu_1 \mu_2},...$ provide a
means for a description of extended objects. If an object is extended,
then not only its center of mass coordinates $x^\mu$, but also the
higher grade coordinates $ x^{\mu_1 \mu_2},~ x^{\mu_1 \mu_2, \mu_3}$,...,
associated with the object extension, are different from zero, in general.
Those higher grade coordinates model the extended object. They are a
generalization of the concept of center of mass \ci{PavsicArena}.

Since $x^M$ assumes any real value, the set of all possible
$X$ forms a $2^n$-dimensional manifold, called {\it Clifford space},
or shortly {\it $C$-space}.

Let us define the quadratic form by means of the scalar product
\be
    |\dd X|^2 \equiv \dd X^{\ddagger} * \dd X = \dd x^M \dd x^N
   G_{MN} \equiv \dd x^M \dd x_M
\lbl{2.7}
\ee
where the metric of $C$-space is given by
\be
      G_{MN} = \gam_M^{\ddagger} * \gam_N
\lbl{2.8}
\ee
The operation $\ddagger$ reverses the order of vectors:
\be
    (\gam_{\mu_1} \gam_{\mu_2}...\gam_{\mu_r})^{\ddagger}
  = \gam_{\mu_r} ... \gam_{\mu_2} \gam_{\mu_1}
\lbl{2.8a}
\ee
Indices are lowered and raised by $G_{MN}$ and its inverse
$G^{MN}$, respectively. The following relation is satisfied:
 \be
     G^{MJ} G_{JN} = {\delta^M}_N
\lbl{2.9}
\ee

Considering the definition (\ref{2.8}) for the $C$-space metric, one could
ask why just that definition, which involves reversion, and not a
slightly different definition, e.g., without reversion.
That reversion is necessary for consistency we can demonstrate by the
following example. Let us take a polyvector which has only the 2-vector
component different from zero:
\be
       x^N = (0,0,x^{\alpha \beta},0,0,...,0)
\lbl{2.5a}
\ee
Then the covariant components are
\be
    x_M = G_{MN} x^N = {1\oo 2} G_{M[\alpha \beta]} x^{\alpha \beta}
\lbl{2.5b}
\ee
Since the metric $G_{MN}$ is block diagonal, so that $G_{M[\alpha \beta]}$
differs from zero only if $M$ is bivector index, we have
\be
    x_M = x_{\mu \nu} = {1\oo 2} G_{[\mu \nu][\alpha \beta]} x^{\alpha
    \beta}
\lbl{2.5c}
\ee

From the definition (\ref{2.8}) we find
\be
    G_{[\mu \nu][\alpha \beta]}= (\gam_\mu \wg \gam_\nu)^{\ddagger}
    *(\gam_\alpha \wg \gam_\beta) = (\gam_\nu \wg \gam_\mu)*
    (\gam_\alpha \wg \gam_\beta) = 
    g_{\mu \alpha} g_{\nu \beta} - g_{\mu \beta} g_{\nu \alpha}
\lbl{2.5d}
\ee
Inserting (\ref{2.5d}) into (\ref{2.5c}) we obtain
\be
    x_{\mu \nu} = {1\oo 2} (g_{\mu \alpha} g_{\nu \beta} - 
    g_{\mu \beta} g_{\nu \alpha}) x^{\alpha \beta} =
    g_{\mu \alpha} g_{\nu \beta} x^{\alpha \beta}
\lbl{2.5e}
\ee
From the fact that the usual metric $g_{\mu \nu}$ lowers the indices
$\mu,~\nu,~\alpha,~\beta,...$, so that
\be
     g_{\mu \alpha} g_{\nu \beta} x^{\alpha \beta} = x_{\mu \nu}
\lbl{2.5f}
\ee
It follows that eq. (\ref{2.5e}) is just an identity.

Had we defined the $C$-space metric without employing the reversion,
then instead of eq.(\ref{2.5d}) and (\ref{2.5e}) we would have
$G_{[\mu \nu][\alpha \beta]}=-(g_{\mu \alpha} g_{\nu \beta} - 
g_{\mu \beta} g_{\nu \alpha})$ and $x_{\mu \nu} = - 
g_{\mu \alpha} g_{\nu \beta} x^{\alpha \beta} = - x_{\mu \nu}$,
which is a contradiction\footnote{By an analogous derivation we find
that the relation $x^M = {G^M}_N x^N = {\delta^M}_N x^N$ holds
if ${G^M}_N = (\gam^M)^{\ddagger} * \gam_N$. The definition
${G^M}_N = \gam^M * \gam_N$ leads to the contradictory equation
$x^M = {G^M}_N x^N = - {\delta^M}_N x^N$.}.

Eq.(\ref{2.7}) is the expression for the {\it line element} in
$C$-space. If $C$-space is generated from the basis vectors
$\gam_\mu$ of spacetime $M_n$ with signature
$(+ - - - - - ...)$, then the signature of $C$-space is
$(+ + + ... - - -...)$, where the number of plus and minus signs is
the same, namely, $2^n/2$. This has some important
consequences that we are going to investigate in the next sections.

We assume that $2^n$-dimensional Clifford space is the
arena in which physics takes place. We can take $n=4$,
so that the spacetime from which we start is just the
4-dimensional Minkowski space $M_4$. The corresponding
Clifford space has then 16 dimensions. In $C$-space the
usual points, lines, surfaces, volumes and 4-volumes are
all described on the same footing and can be transformed
into each other by rotations in $C$-space (called polydimensional
rotations):
\be
    x'^M = {L^M}_N x^N
\lbl{2.10}
\ee
subjected to the condition $|\dd X'|^2 = |\dd X|^2$.

We can now envisage that physical objects are described
in terms of $x^M = (\sigma, x^\mu, x^{\mu \nu},...)$. The
first straightforward possibility is to introduce a single
parameter $\tau$ and consider a mapping
\be
    \tau \rightarrow x^M = X^M (\tau)
\lbl{2.10a}
\ee
where $X^M (\tau)$ are 16 embedding functions that
describe a world-line in $C$-space. From the point of view of
$C$-space, $X^M (\tau)$ describe a wordlline of a ``point
particle": At every value of $\tau$ we have a {\it point} in
$C$-space. But from the perspective of the underlying
4-dimensional spacetime, $X^M (\tau)$ describe an extended
object, sampled by the center of mass coordinates $X^\mu (\tau)$
and the coordinates
$X^{\mu_1 \mu_2}(\tau),..., X^{\mu_1 \mu_2 \mu_3 \mu_4} (\tau)$.
They are a generalization of the center of mass coordinates in the sense
that they provide information about the object 2-vector, 3-vector, and
4-vector extension and orientation\footnote{A systematic and detailed
treatment is in ref. \ci{PavsicArena}.}

Let the dynamics of such an object be determined by the action
\be
    I[X] = M \int \dd \tau \, ({\dot X}^{\ddagger} *{\dot X})^{1\oo 2} =
   M \int \dd \tau ({\dot X}^M {\dot X}_M)^{1\oo 2}
\lbl{2.11}
\ee
The dynamical variables are given by the polyvector
\be
   X = X^M \gam_M = \sigma {\underline 1} + X^\mu \gam_\mu +
  X^{\mu_1 \mu_2} \gam_{\mu_1 \mu_2} + ... X^{\mu_1 ...\mu_n} 
  \gam_{\mu_1 ... \mu_n}
\lbl{2.11a}
\ee
whilst
\be 
{\dot X} = {\dot X}^M \gam_M = {\dot \sigma} {\underline 1} + 
{\dot X}^\mu \gam_\mu +
  {\dot X}^{\mu_1 \mu_2} \gam_{\mu_1 \mu_2} + ... 
 {\dot X}^{\mu_1 ...\mu_n} \gam_{\mu_1 ... \mu_n}
\lbl{2.11b}
\ee
is the velocity polyvector, where ${\dot X}^M \equiv \dd X^M/\dd \tau$.

In the action (\ref{2.11}) we have a straightforward generalization of the
relativistic point particle in $M_4$:
\be
   I[X^\mu] = m \int \dd \tau ({\dot X}^\mu {\dot X}_\mu)^{1\oo 2}
  \; , \quad \mu = 0,1,2,3
\lbl{2.12}
\ee
 If a particle is extended, then the latter action (\ref{2.12})
provides only a very incomplete
description. A more complete description is given by the action
(\ref{2.11}), in which the $C$-space embedding functions $X^M (\tau)$
sample the objects extension \ci{PavsicArena}.

\section{A toy model: Harmonic oscillator in pseudo-Eucidean space}

\subsection{A simple model in $M_{1,-1}$}

Suppose that instead of usual Harmonic oscillator we have a system given
by the Lagrangian \ci{PseudoHarm}
\be
    L = {1\oo 2} ({\dot x}^2 - {\dot y}^2) - {1\oo 2} \omega^2
       (x^2 - y^2)
\lbl{3.1}
\ee
If we derive the equations of motion we find that they are indistinguishable
form those of the usual harmonic oscillator:
\be
    {\ddot x} + \omega^2 x = 0 \; , \qquad {\ddot y} + \omega^2 y = 0
\lbl{3.1a}
\ee
The change of sign in front of the $y$-term have no influence on the
equations of motion.

However, a difference occurs when we calculate the {\it canonical momenta}
\be
    p_x = {{\p L}\oo {\p {\dot x}}} = {\dot x} \; , \qquad
    p_y = {{\p L}\oo {\p {\dot y}}} = - {\dot y}
\lbl{3.2}
\ee
and the {\it Hamiltonian}
\be
    H = p_x {\dot x} + p_y {\dot y} - L = {1\oo 2} (p_x^2 - p_y^2) +
    {\omega^2\oo 2} (x^2 - y^2)
\lbl{3.3}
\ee
We see that the $y$-terms have negative contribution to the energy of our
system, and that energy has no lower bound (as well as no upper bound).
For a usual physical system this indicates its instability. But for
a system derived from the Lagrangian (\ref{3.1}) this is not the case.
In eq.(\ref{3.1}) the kinetic term for the $y$-component has negative
sign, whilst that for the $x$-component has positive sign. Therefore,
the equations of motion are
\be
    {\ddot x} = - {{\p V}\oo {\p x}} \; , \qquad 
    {\ddot y} =  {{\p V}\oo {\p y}}
\lbl{3.4}
\ee
where $V= {1\oo 2} \omega^2 (x^2 - y^2)$ is the potential. For the
$x$-component the force is given by {\it minus} gradient of the potential,
whilst for the $y$-component the force is given by {\it plus} gradient
of the potential. Therefore, the criterium for the stability of motion for
the $y$-degree of freedom is that the potential has to have a maximum
in the $(y,V)$-plane. This is just opposite to the case of the $x$-degree
of freedom where stability requires a minimum of $V(x,y)$ in the plane
$(x,V)$.

This was just a more sophisticated explanation which involves the concept
of energy. That our system is indeed stable can be directly read from the
equations of motion (\ref{3.2}) from which it follows that both
$x$ and $y$ degrees of freedom oscillate around the origin $x=0, ~y=0$.

In the Hamiltonian form the equations of motion read
\bear
  {\dot x} &=& \lbrace x,H \rbrace = {{\p H}\oo {\p p_x}} = p_x \nonumber \\
  {\dot y} &=& \lbrace y,H \rbrace = {{\p H}\oo {\p p_y}} = - p_y \nonumber \\
  {\dot p}_x &=& \lbrace p_x, H \rbrace = -{{\p H}\oo {\p p_x}} =
  - \omega^2 x \nonumber \\
  {\dot p}_y &=& \lbrace p_y, H \rbrace = -{{\p H}\oo {\p p_y}} =
   \omega^2 y   \lbl{3.5}
\ear
where the Poisson brackets are defined as usual. In particular we have
\be
   \lbrace x, p_x \rbrace = 1 \; , \qquad \lbrace y, p_y \rbrace = 1
\lbl{3.6}
\ee

The system can be quantized by replacing the canonically conjugate variables
$(x,p_x)$ and $(y,p_y)$ by operators satisfying the following commutation
relations\footnote{We use units in which $\hbar =c=1$.}:
\be
     [x,p_x] =i \; , \qquad  [y,p_y] = i
\lbl{3.7}
\ee

Let us introduce non Hermitian operators
\be
     c_x = {1\oo \sqrt{2}} (\sqrt{\omega} \, x + {i\oo {\sqrt{\omega}}} \, p_x )
     \; , \quad 
     c_x^{\dagger}
      = {1\oo \sqrt{2}} (\sqrt{\omega} \, x - {i\oo {\sqrt{\omega}}} \, p_x )
\lbl{3.8a}
\ee
\be
c_y = {1\oo \sqrt{2}} (\sqrt{\omega} \, y + {i\oo {\sqrt{\omega}}} \, p_y )
     \; , \quad 
     c_y^{\dagger}
      = {1\oo \sqrt{2}} (\sqrt{\omega} \, y - {i\oo {\sqrt{\omega}}} \, p_y )
\lbl{3.8b}
\ee
They satisfy the commutation relations
\be
    [c_x,c_x^{\dagger}] = 1 \; , \qquad [c_y,c_y^{\dagger}] = 1
\lbl{3.9}
\ee
\be
      [c_x,c_y] = [c_x^{\dagger},c_y^{\dagger}] = 0
\lbl{3.10}
\ee
In terms of the new variables (\ref{3.8a}),(\ref{3.8b}) the Hamiltonian reads
\be
    H = {1\oo 2} \omega (c_x^{\dagger} c_x + c_x c_x^{\dg} - c_y^{\dg} c_y
    - c_y c_y^{\dagger})
\lbl{3.11}
\ee

Let us define vacuum state according to
\be
    c_x |0 \rangle = 0 \; , \qquad c_y |0 \rangle = 0
\lbl{3.12}
\ee
so that $c_x,~c_y$ are annihilation and  $c_x^{\dg},~c_y^{\dg}$ creation
operators. Using (\ref{3.9}) we find
\be
    H = \omega (c_x^{\dg} c_x - c_y^{\dg} c_y)
\lbl{3.13}
\ee
In the latter expression we have ordered the operators so that creation
operators are on the left and annihilation operators on the right.
{\it We see that zero point energy in the Hamiltonian (\ref{3.13}) cancels
out!}

It is instructive to consider the $(x,y)$-representation in which the momentum
operators are $p_x = - i \p/\p x$, $p_y = - i \p/\p y$, and writing
$\langle x,y|0 \rangle \equiv \psi_0 (x,y)$. Eqs. (\ref{3.12}) then
become
\be
      {1\oo 2} \left ( \sqrt{\omega} x + {1\oo \sqrt{\omega}} \,
      {\p \oo {\p x}} \right ) \psi_0 (x,y) = 0 \lbl{3.14a}
\ee
\be
      {1\oo 2} \left ( \sqrt{\omega} y  + {1\oo \sqrt{\omega}} \,
      {\p \oo {\p y}} \right ) \psi_0 (x,y) = 0 \lbl{3.14b}
\ee
A solution which is in agreement with the probability interpretation reads
\be
    \psi_0 = {2 \pi \oo \omega} \, {\rm e}^{-{1\oo 2} \omega (x^2+y^2)}
\lbl{3.15}
\ee
and is normalized according to $\int \psi_0^2 \, \dd x \, \dd y = 1$.
A particle described by the wave function $\psi_0$ of eq. (\ref{3.15}) is
localized around the origin. The excited states obtained by applying
the product of operators $c_x^\dg$, $c_y^\dg$ to the vacuum state are
also localized. This is in agreement with the property that the corresponding
classical particle moving according to the equations of motion
(\ref{3.2}) is also ``localized" in the vicinity of the origin.

All states $|\psi \rangle$ of our system have positive norm. For
instance
\be
    \langle 0|c c^\dg|0 \rangle = \langle 0|[c,c^\dg]|0 \rangle =
    \langle 0|0 \rangle = \int \psi^2 \, \dd x \, \dd y = 1
\lbl{3.16}
\ee
This is so because of our choice of vacuum (\ref{3.12}). Had we defined vacuum
differently, e.g., by $c_x |0 \rangle = 0$, $c_y^\dg |0 \rangle = 0$
we would have states with negative norm in our theory.

Our action (\ref{3.1}) is invariant under the pseudo rotations in 
$M_{1,-1}$:
\be
     x' = {{x-v y}\oo {\sqrt{1-v^2}}} \; , \qquad 
     y' = {{y-v x}\oo {\sqrt{1-v^2}}}
\lbl{3.17}
\ee
where $v$ is the parameter of the transformation. In a new reference frame
$S'$ we have $c_{x'}|0 \rangle = 0$, $c_{y'}|0 \rangle = 0$, and a
normalized solution is
\be
     \psi_0' = {{2 \pi}\oo \omega} \, {\rm e}^{- {\omega\oo 2}(x'^2 +y'^2)}
\lbl{3.18}
\ee
Expressed in terms of the old coordinates the latter wave function reads
\be
     \psi_0' = {{2 \pi}\oo \omega} \, {\rm exp} \left [- {\omega \oo 2}
     (x^2+y^2){{1+v^2}\oo{1-v^2}} + {{2 \omega v}\oo {1-v^2}}\,x y \right ]
\lbl{3.19}
\ee
Let us now decide to observe everything from a fixed frame $S$.
It is not difficult to find out that $\psi_0'$ of eq. (\ref{3.19})
satisfies the Schr\" odinger equation in the old frame $S$:
\be
   -{1\oo 2} \left ( {{\p^2 \psi_0'}\oo {\p x^2}} - {\p^2 \psi_0'\oo 
   \p y^2} \right ) + {\omega^2 \oo 2} (x^2-y^2) \psi_0' = 0
\lbl{3.20}
\ee
In a given reference frame we have thus a family of solutions
\be
   \psi_0 (x,y;v) = {{2 \pi}\oo \omega} \, {\rm exp} \left [- {\omega \oo 2}
     (x^2+y^2){{1+v^2}\oo{1-v^2}} + {{2 \omega v}\oo {1-v^2}}\,x y \right ]
\lbl{3.21}
\ee
{\it all having zero energy}. For $v=0$ we recover eq. (\ref{3.15}).

The first excited state in the $x$ and $y$ direction, respectively, are
\be
    \psi_{10} ={1\oo \sqrt{2}} \left ( \sqrt{\omega} x - {1\oo \sqrt{\omega}} \,
    {\p \oo {\p x}} \right ) \psi_0 = \sqrt{2} \sqrt{\omega} \, x \,
    {\rm e}^{-{1\oo 2} \omega (x^2+y^2)}
\lbl{3.21a}
\ee
\be
    \psi_{01} ={1\oo \sqrt{2}} \left ( \sqrt{\omega} y - {1\oo \sqrt{\omega}} \,
    {\p \oo {\p y}} \right ) \psi_0 = \sqrt{2} \sqrt{\omega} \, y \,
    {\rm e}^{-{1\oo 2} \omega (x^2+y^2)}
\lbl{3.21b}
\ee
In a new reference frame $S'$ obtained by the transformation (\ref{3.17})
the vacuum state is given by (\ref{3.18}) and the excited states are
given by the same equations (\ref{3.21a}),(\ref{3.21b}) in which $x$ and $y$ are
replaced by $x'$ and $y'$. Expressing $x'$ and $y'$ in terms of
$x$ and $y$ by using the transformation (\ref{3.17}) we obtain
\be
   \psi_{10} (x,y;v) = {{\sqrt{2 \omega}}\oo \sqrt{1-v^2}} \, (x- v y)
   {\rm exp} \left [ -{\omega\oo 2} \left ( {{(x-v y)^2}\oo {1-v^2}} 
   + {{(y-v x)^2}\oo {1-v^2}} \right ) \right ]
\lbl{3.21c}
\ee
\be
   \psi_{01} (x,y;v) = {{\sqrt{2 \omega}}\oo \sqrt{1-v^2}} \, (y- v x)
   {\rm exp} \left [ -{\omega\oo 2} \left ( {{(x-v y)^2}\oo {1-v^2}} 
   + {{(y-v x)^2}\oo {1-v^2}} \right ) \right ]
\lbl{3.21d}
\ee
which are now the states as observed from the reference frame $S$.
A state $\psi(x,y;v)$ is obtained from the state $\psi(x,y;0)$ by an
{\it active} pseudo rotation of the form (\ref{3.17}).

One can verify that  the states $\psi_{10} (x,y;v)$ and $\psi_{01} (x,y;v)$
satisfy the Schr\" odinger equation in frame S
for arbitrary value of the parameter $v$:
\be
    \left [ -{1\oo 2} \left ( {\p^2 \oo {\p x^2}} + {\p^2 \oo {\p y^2}} 
    \right ) + {\omega\oo 2} \, (x^2-y^2) \right ] \psi_{10} (x,y;v)
    = \omega \, \psi_{10} (x,y;v)
\lbl{3.21e}
\ee
\be
    \left [ -{1\oo 2} \left ( {\p^2 \oo {\p x^2}} + {\p^2 \oo {\p y^2}} 
    \right ) + {\omega\oo 2} \, (x^2-y^2) \right ] \psi_{01} (x,y;v)
    = - \omega \, \psi_{01} (x,y;v)
\lbl{3.21f}
\ee
The state $\psi_{10} (x,y;v)$ has energy $E=\omega$, whilst the state
$\psi_{01} (x,y;v)$ has energy $E=- \omega$. A generic excited state
$\psi_{nm} (x,y;v)$ has energy $E= \omega (n-m)$. Since $v$ is an arbitrary
parameter $v \in [0,1]$ we have for fixed $m,n$ a family of states
$\lbrace \psi_{mn} (x,y;v) \rbrace $ all having the same energy
$E= \omega (n-m)$. Not only the states (\ref{3.21a}),(\ref{3.21b}), but
also the states (\ref{3.21c}),(\ref{3.21d}) and the higher excited states
for arbitrary values of $v$ satisfy the same Schr\" odinger equation.

A particular model of relativistic Harmonic oscillator in spacetime
$M_{1,-3}$ has been considered in refs. \ci{Kim} with a motivation
to explain partons.

\subsection{Generalization to $M_{r,-s}$}

Let us now consider the harmonic oscillator in a pseudo-Euclidean space
of arbitrary dimensiona and signature. Instead of (\ref{3.1}) we have
now the Lagrangian
\be
    L= {1\oo 2} {\dot x}^\mu {\dot x}_\mu - {1\oo 2} \omega^2 x^\mu x_\mu
\lbl{3.22}
\ee
The canonical momenta are
\be
    p_\mu = {{\p L}\oo {\p {\dot x}^\mu}} = {\dot x}_\mu = \eta_{\mu \nu}
    {\dot x}^\nu
\lbl{3.23}
\ee
The Hamiltonia is
\be
    H = {1\oo 2} p^\mu p_\mu + {1\oo 2} \omega^2 x^\mu x_\mu
\lbl{3.24}
\ee

Upon quantization the phase space variables $(x^\mu, p_\nu)$ become
operators satisfying
\be
    [x^\mu,p_\nu] = i {\delta^\mu}_\nu \quad {\rm or} \quad 
    [x^\mu,p^\nu] = i \eta^{\mu \nu}
\lbl{3.25}
\ee
A straightforward generalization of the non Hermitian operators 
(\ref{3.8a}),(\ref{3.8b}) is
\be
    c^\mu = {1\oo \sqrt{2}} \left ( \sqrt{\omega} x^\mu + {i \oo \sqrt{\omega}}
    \, p_\mu \right )\; , \quad 
\lbl{3.26a}
\ee       
\be
{c^\mu}^\dg = {1\oo \sqrt{2}} \left ( \sqrt{\omega} x^\mu - {i \oo \sqrt{\omega}}
    \, p_\mu \right )
\lbl{3.26b}
\ee
Notice that in the definition of $c^\mu$, ${c^\mu}^\dg$ we take 
{\it contravariant} components of coordinates and {\it covariant} components
of $p_\mu$. This is in agreement with the definition 
(\ref{3.8a}),(\ref{3.8b}) where
$p_x$, $p_y$ defined in eq. (\ref{3.2}) are in fact the covariant
components. Using (\ref{3.26a}),(\ref{3.26b}), the Hamiltonian operator
(\ref{3.24}) becomes
\be
    H = {1\oo 2} \, \omega (c_\mu^\dg c^\mu + c^\mu c_\mu^\dg )
\lbl{3.27}
\ee

The definition of vacuum state which is a generalization of the definition
(\ref{3.12}) reads
\be
     c^\mu |0 \rangle = 0
\lbl{3.28}
\ee

The annihilation and creation operators $c^\mu$, ${c^\nu}^\dg$
satisfy the commutation relations
\be
        [c^\mu,{c^\nu}^\dg] = \delta^{\mu \nu}
\lbl{3.29}
\ee
where $\delta^{\mu \nu}$ is the Kronecker symbol (having $+1$ values on
the diagonal and zero elsewhere). Using (\ref{3.29}) we have
\be
     c^\mu c_\mu^\dg = \eta_{\mu \nu} c^\mu c^{\nu \dg} =
     \eta_{\mu \nu} (c^{\nu \dg} c^\mu + \delta^{\mu \nu} )
     \equiv c^{\mu \dg} c_\mu + r - s
\lbl{3.30}
\ee
where we have written $\eta_{\mu \nu} c^{\nu \dg} c^\mu =
\eta_{\mu \nu} c^{\mu \dg} c^\nu \equiv c^{\mu \dg} c_\mu$ and
$\eta_{\mu \nu} \delta^{\mu \nu} = r-s$. Here $r$ is the number of positive
and $s$ negative signature components.

The Hamiltonian is thus
\be
     H = \omega (c_\mu^\dg c^\mu + {r\oo 2} - {s\oo 2} )
\lbl{3.31}
\ee
In particular, if $r=s$, the zero point energies cancel out.

The definition (\ref{3.26a}),(\ref{3.26b}) of the creation and annihilation
operators is
natural, because it takes a superposition of coordinates and canonical
momenta. But from the viewpoint of the tensor calculus the notation in
eq, (\ref{3.26a}),(\ref{3.26b} is not very fortunate,
because normally we do not sum
covariant and contravariant components. Therefore I will now rewrite the
theory by using the usual formalism in which creation and annihilation
operators are defined in terms of contravariant components $x^\mu$ and
$p^\mu$:
\be
     a^\mu = {1\oo 2} \left ( \sqrt{\omega}\, x^\mu + {i\oo \sqrt{\omega}}
     \, p^\mu \right )
\lbl{3.32}
\ee
\be
     a^{\mu \dg}= {1\oo 2} \left ( \sqrt{\omega}\, x^\mu - {i\oo \sqrt{\omega}}
     \, p^\mu \right )
\lbl{3.33}
\ee     
The non vanishing commutators are
\be
    [a^\mu,a_\nu^\dg ] = {\delta^\mu}_\nu \quad {\rm or} \quad
    [a^\mu,a^{\nu \dg} ] = \eta^{\mu \nu}
\lbl{3.34}
\ee
Hamiltonian is
\be
    H = {1\oo 2} \omega \, (a^{\mu \dg} a_\mu + a_\mu a^{\mu \dg} )
\lbl{3.35}
\ee
Given the operators (\ref{3.32}),(\ref{3.33}), let us consider two possible
definitions of vacuum.

\vs{2mm}

{\it $1^{st}$ possible definition of vacuum}

This is the definition that is usually adopted and it reads
\be
    a^\mu |0 \rangle = 0
\lbl{3.36}
\ee
The Hamiltonian, after using (\ref{3.34}) and (\ref{3.36}),
is
\be
    H = \omega \left ( a^{\mu \dg} a_\mu + {d\oo 2} \right )
\lbl{3.37}
\ee
where $d= \eta^{\mu \nu} \eta_{\mu \nu} = r+s$ is the dimension of $M_{r,-s}$.
There occurs the non vanishing zero point energy. The eigenstates of $H$ are
all positive. This is so even for those terms in $H$ which belong to
negative signature: negative sign of a term in $a^{\mu \dg} a_\mu$ is
compensated by negative sign in the commutation relations (\ref{3.34}).
Also the expectation values between the eigenstates $|A \rangle$ of
$H$ calculated according to
\be
    \langle H \rangle = {{\langle A|H|A \rangle }\oo {\langle A|A \rangle}}
\lbl{3.38}
\ee
are always positive, since the negative norm in the denominator and
negative norm in the numerator together give 1.

\vs{2mm}

{\it  $2^{nd}$ possible definition of vacuum}

Let us split $a^{\mu} = (a^{\alpha}, 
a^{\bar \alpha})$, where indices
$\alpha$, ${\bar \alpha}$ refer to the components with positive and
negative signature, respectively, and define the vacuum according to
\ci{Jackiw} (see also \ci{PseudoHarm}
\be
    a^{\alpha} |0 \rangle = 0  \; \qquad \qquad a^{{\bar \alpha} \dagger}
    |0 \rangle = 0
\lbl{3.39}
\ee

\nnn Using (\ref{3.34}) we obtain the Hamiltonian in which the annihilation
operators, defined according to eq.\,(\ref{3.39}), are on the right:
\be
      H = \omega \left ( a^{\alpha \dagger} a_{\alpha} + {r \oo 2} +
           a_{\bar \alpha} a^{{\bar \alpha} \dagger} - {s \oo 2} \right )
\lbl{3.40}
\ee

\nnn where ${\delta_{\alpha}}^{\alpha} = r$ and ${{\delta}_{\bar \alpha}}^
{\bar \alpha} = s$. If the number of positive and negative signature
components is the same, i.e., $r = s$, then the Hamiltonian (\ref{3.40})
has vanishing zero-point energy:
\be
H = \omega (a^{\alpha \dagger} a_{\alpha} + 
           a_{\bar \alpha} a^{{\bar \alpha} \dagger})
\lbl{3.41}
\ee

\nnn Its eigenvalues are positive or negative, depending on which component
(positive or negative signature) are excited. In $x$-representation
the vacuum state (\ref{3.39}) is
\be
     \psi_0 = {\left ( {{2 \pi} \oo \omega} \right ) }^{d/2} {\rm exp}[-{\omega
     \oo 2} \delta_{\mu \nu} x^{\mu} x^{\nu}]
\lbl{3.42}
\ee

\nnn where the Kronecker symbol $\delta_{\mu \nu}$ has values +1 for $\mu =\nu$
and 0 otherwise.
It is a solution of the Schr\" odinger equation
\be
-(1/2) \p^{\mu} \p_{\mu} \psi_0 + (\omega^2/2) x^{\mu} x_{\mu} \psi_0
= E_0 \psi_0 
\lbl{3.43}
\ee
with $E_0 = \omega ({1\oo 2} + {1\oo 2} + .... - {1\oo 2} -
{1\oo 2}- ...)$.
One can also easily verify that there are no negative norm states.

Let us now consider a pseudo rotation 
\be
    x'^\mu = {L^\mu}_\nu x^\nu
\lbl{3.A0}
\ee
The transformed vacuum wave function reads (see Sec.(3.1) for specific examples
in $M_{1,-1}$):
\be
   \psi_0 (x') = {\rm exp} \left [ -{\omega\oo 2} \, 
   \delta_{\mu \nu} x'^\mu x'^\nu \right ] = 
   {\rm exp} \left [ -{\omega\oo2} \delta_{\mu \nu} {L^\mu}_\rho {L^\nu}_\sigma
   x^\rho x^\sigma \right ] = \psi'(x)
\lbl{3.A1}
\ee
In order to show that also $\psi'(x)$ is a solution of the Schr\"odinger
equation we use
\be
   \p_\beta \psi'_0 = - \omega \delta_{\mu \nu} {L^\mu}_\beta {L^\nu}_\sigma
   x^\sigma \psi'
\lbl{3.A2}
\ee
and
\bear
     \eta^{\alpha \beta} \p_\alpha \p_\beta \psi'_0 (x) &=& \left (
     - \omega \eta^{\alpha \beta} {L^\mu}_\alpha {L^\nu}_\beta \delta_{\mu \nu}
     + \omega^2 \eta^{\alpha \beta} \delta_{\mu \nu} {L^\mu}_\beta
     {L^\nu}_\sigma x^\sigma {L^\epsilon}_\alpha {L^\gam}_\rho x^\rho
     \delta_{\epsilon \gam} \right ) \psi'_0 (x) \nonumber \\
      &=& \omega^2 x^\alpha
     x_\alpha \psi'_0 (x)
\lbl{3.A3}
\ear
where we have used $\eta^{\alpha \beta} {L^\mu}_\beta {L^\nu}_\alpha
= \eta^{\mu \nu}$ and $\eta^{\mu \nu} \delta_{\mu \nu} = r-s=0$. From
eq.(\ref{3.A3}) it follows that $\psi'_0 (x)$ satisfies 
\be
    -{1\oo 2} \eta^{\alpha \beta} \, \p_\alpha \p_\beta \psi'_0 (x) +
    {\omega^2\oo 2} \, x^\alpha x_\alpha \psi'_0 (x) = 0
\lbl{3.A4}
\ee
which is the Schr\" odinger equation for the (actively) transformed vacuum
wave function. Hence the theory is {\it covariant}, although the vacuum,
defined according to eq.\,(\ref{3.39}), is not {\it invariant} under the
pseudo rotations. 

In general, for the excited states, let us start from the Schr\" odinger equation
in a reference frame $S'$:
\be
   -{1\oo 2} \eta^{\rho \sigma}\, \p'_\rho \p'_\sigma \psi (x')
   {\omega^2\oo 2} \, x'^\mu x'_\mu \psi (x') = E \psi (x')
\lbl{3.A5}
\ee
Let us now apply to eq.(\ref{3.A5}) the pseudo rotation 
$x'^\mu = {L^\mu}_\rho x^\rho$, briefly, $x' = L(x)$:
\be
    -{1\oo 2} \eta^{\rho \sigma} \, {L^\mu}_\rho {L^\nu}_\sigma
    \p_\mu \p_\nu \psi (L(x)) +
    {\omega^2\oo 2} \, x^\mu x_\mu \psi (L(x))) = E \psi (L(x))
\lbl{3.A6}
\ee
Writing $\psi(L(x)) = \psi'(x)$ we find
\be
    -{1\oo 2} \p^\mu \p_\mu \psi' (x) + {\omega^2\oo 2} x^\mu x_\mu \,
    \psi' (x) = E \psi'(x)
\lbl{3.A7}
\ee
which means that the (actively) transformed excited state $\psi'(x)$ (as
observed from the frame $S$) satisfied the Schr\" odinger equation. So
we have found that $\psi(x)$ as well as $\psi'(x)$ are solutions of
the Schr\" odinger equation in $S$ and they both have the same energy $E$.
In general, in a given reference frame we have thus a degeneracy of
solutions with the same energy (see also ref. \ci{Kim}).

\section{Quantum field theory}

\subsection{A system of scalar fields}

Let us now turn to the theory of $n$ scalar fields $\phi^a$,
$a=0,1,2,...,n-1$, over the 4-dimensional spacetime parametrized by
coordinates $x^\mu$, $\mu = 0,1,2,3$. The action for such system is
\be
    I[\phi^a] = {1\oo 2} \int \dd^4 x \, \sqrt{-g} (g^{\mu \nu}
    \p_\mu \phi^a \, \p_\nu \phi^b - m^2 \phi^a \phi^b ) \gamma_{ab}
\lbl{4.1}
\ee
where $\gam_{ab}$ is the metric in the space of fields $\phi^a$,
and $g_{\mu \nu}$ the metric of spacetime. Let us assume that $g_{\mu \nu}
=\eta_{\mu \nu}$ is the metric of flat spacetime.

The canonical momenta are
\be
     \pi_a = {\p {\cal L}\oo \p \p_0 \phi^a} = \p^0 \phi_a = \p_0 \phi_a \equiv
     {\dot \phi}_a
\lbl{4.2}
\ee

Upon quantization the following equal time commutation relations are
satisfied:
\be
    [\phi^a ({\bf x}),\pi_b ({\bf x}')] = i \delta^3 ({\bf x} - {\bf x}')
    {\delta^a}_b
\lbl{4.3}
\ee
The Hamiltonian is given by
\be
    H= {1\oo 2} \int \dd^3 x \, ( {\dot \phi}^a {\dot \phi}^b -
    \p_i \phi^a \p^i \phi^b + m^2 \phi^a \phi^b) \gam_{ab}
\lbl{4.4}
\ee
where $i= 1,2,...,n-1$. We shall assume that $\gam_{ab}$ is diagonal.
A general solution to the equations of motion derived from the action
(\ref{4.1}) can be written in the form
\be
    \phi^a = \int {{\dd^3 {\bf k}}\oo {(2 \pi)^3}} \, 
    {1\oo {2 \omega_{\bf k}}} (a^a ({\bf k}) {\rm e}^{-ikx} + {a^a}^\dagger
    ({\bf k}) {\rm e}^{ikx} )
\lbl{4.5}
\ee
Here\footnote{We use units in which $\hbar = c = 1$.} $\omega_{\bf k} 
\equiv |\sqrt{m^2 + {\bf k}^2}|$. The creation and annihilation operators
satisfy the commutation relations
\be
    [a^a ({\bf k}),{a_b}^\dagger ({\bf k'})] = (2 \pi)^3 \, 2 \omega_{\bf k}
    \, \delta^3 ({\bf k} - {\bf k}') {\delta^a}_b
\lbl{4.5a}
\ee
or
\be
    [a^a ({\bf k}),{a^b}^\dagger ({\bf k'})] = (2 \pi)^3 \, 2 \omega_{\bf k}
    \, \delta^3 ({\bf k} - {\bf k}') \gam^{ab}
\lbl{4.5b}
\ee   
Inserting the expansion (\ref{4.5}) of fields $\phi^a$ into the Hamiltonian
(\ref{4.4}) we obtain
\be
    H = {1\oo 2} \int {{\dd^3 {\bf k}}\oo {(2 \pi)^3}} \, 
    {\omega_{\bf k}\oo {2 \omega_{\bf k}}} ({a^a}^\dagger ({\bf k})
    a^b ({\bf k}) + a^a ({\bf k}) {a^b}^\dagger ({\bf k}) ) \gam_{ab}
\lbl{4.6}
\ee

Let us assume that the signature of the metric $\gam_{ab}$ is pseudo-Euclidean,
and let us write
\be
    a^a ({\bf k}) = (a^\alpha, a^{\bar \alpha})
\lbl{4.7}
\ee
where $\alpha$ denotes positive and ${\bar \alpha}$ negative signature
components.

We will define vacuum according to
\be
    a^\alpha ({\bf k}) |0 \rangle = 0 \; , \quad {a^{\bar \alpha}}^\dagger ({\bf k})
    |0 \rangle = 0
\lbl{4.8}
\ee
If we order the operators so that the annihilation operators, defined in 
eq.\,(\ref{4.8}), are on the right, and use the commutation relations
(\ref{4.5b}), we find

\be
    H = \int {{\dd^3 {\bf k}}\oo {(2 \pi)^3}} \, 
    {\omega_{\bf k}\oo {2 \omega_{\bf k}}} ({a^\alpha}^\dagger ({\bf k})
    a_\alpha ({\bf k}) + a^{\bar \alpha} ({\bf k}) 
    {a_{\bar \alpha}}^\dagger) + {1\oo 2} \int \dd^3 {\bf k} \, \omega_{\bf k}
    \delta^3 (0) (r - s)
\lbl{4.9}
\ee
where $r={\delta^\alpha}_\alpha$ and $s= {\delta^{\bar \alpha}}_{\bar \alpha}$.
In the case in which the signature has equal number of plus and minus signs,
i.e., when $r=s$, the zero point energies cancel out from the
Hamiltonian.

\subsection{Genaralization to Clifford space}

Now a question arises as to why should the space of fields have the metric
with $r=s$. Isn't it an ad hoc assumption? The answer is as follows.
We can consider the space of fields $V_n$ just as a starting space,
with basis $e_a,~a=0,1,2,...,n-1$, from which we generate the
$2^n$-dimensional {\it Clifford space} ${\cal C}_{V_n}$ with basis
$e_A = ({\underline 1},e_a, e_{a_1 a_2},...,e_{a_1...a_n}),~a_1<a_2<...<a_n$.
If $V_n$ is a Euclidean space so that $e_a \cdot e_b = \delta_{ab}$ is
the Euclidean metric, then also the metric ${e_A}^\dagger *e_B$ of
${\cal C}_{V_n}$ is Eucliddean. But, as it was pointed out in refs. 
\ci{Pavsic,PavsicBook},
instead of the basis $e_A$ we can take another basis, e.g.,
\be
    \gamma_A = ({\underline 1},\gam_a,\gam_{a_1a_2},...,\gam_{a_1...a_n})
\lbl{4.10}
\ee
generated from the set of Clifford numbers $\gam_a = (e_0,e_i e_0),~
a=0,1,2,...,n-1;~i=1,2,...,n$ satisfying
\be
    \gam_a \cdot \gam_b \equiv {1\oo 2} (\gam_a \gam_b + \gam_b \gam_a) = 
    \eta_{ab}
\lbl{4.11}
\ee
The metric
\be     {\gam_A}^\ddagger * \gam_B = G_{AB}
\lbl{4.13}
\ee
defined with respect to the new basis is pseudo-Euclidean, its signature
having $2^n/2$ plus and $2^n/2$ minus signs.

We assume that a field theory should be formulated in $C$-space in which the
metric is given by eq.(\ref{4.13}). Instead
of the action (\ref{4.1}) we thus consider its generalization to $C$-space:
\be
     I = {1\oo 2} \int \dd^4 x \, \sqrt{-g} (g^{\mu \nu}
     \p_\mu \phi^A \p_\nu \phi^B - m^2 \phi^A \phi^B) G_{AB}
\lbl{4.14}
\ee
Here $\phi^A \gam_A$ is a polyvector field. Since the metric $G_{AB}$ has
signature $(+++ ... - - - ...) = (R+, S-)$ with $R=S$, zero point energies
of a system based on the action (\ref{4.14}) cancel out: vacuum energy
vanishes. Consequently, in such a theory there is no cosmological problem
\ci{PseudoHarm}. The small cosmological constant, as recently observed, could be
a residual effect of something else.

Cancellation of vacuum energies in the theory does not exclude \ci{PseudoHarm}
the existence of well known effects, such as Casimir effect, which is 
a manifestation of vacuum energies.
\vs{5mm}

{\bf 3. Strings and Clifford space}

\vs{2mm}

Usual strings are described by the mapping $(\tau,\sigma) \rightarrow
x^\mu = X^\mu (\tau, \sigma)$, where the embedding functions
$X^\mu (\tau,\sigma)$ describe a 2-dimensional worldsheet swept
by a string. The action is given by the requirement that the area of the
worldsheet be ``minimal" (extremal). Such action is invariant under
reparametrizations of $(\tau,\sigma)$. There are several equivalent
forms of the action including the ``$\sigma$-model action" which, in
the conformal gauge, can be written as
\be
   I[X^\mu] = {\kappa\oo 2} \int \dd \tau \, \dd \sigma \, ({\dot X}^\mu
   {\dot X}_\mu - X'^\mu X'_\mu)
\lbl{4A.1}
\ee
where ${\dot X}^\mu \equiv \dd X^\mu/\dd \tau$ and $X'^\mu \equiv
\dd X^\mu/\dd \sigma$. Here $\kappa$ is the string tension, usually
written as $\kappa = 1/(2 \pi \alpha')$.

String coordinates $X^\mu$ and momenta $P_\mu = \p L/\p {\dot X}^\mu
=\kappa {\dot X}_\mu$
satisfy the following constraints $(\sigma \in [0,\pi])$:
\be
      \varphi_1 (\sigma) = P^\mu P_\mu + {{X'^\mu X'_\mu}\oo {(2 \pi 
  \alpha')^2}} \approx 0 \; \qquad \varphi_2 (\sigma) =
   {{P^\mu X'_\mu}\oo {\pi \alpha'}} \approx 0
\lbl{4A.1a}
\ee
which can be written as a single constraint on the interval $\sigma \in
[- \pi, \pi]$
\be
     \Pi^\mu \Pi_\mu (\sigma) \approx 0 \; \qquad 
     \Pi^\mu = P^\mu + {X'^\mu\oo 2 \pi \alpha'}
\lbl{4A.1b}
\ee     
to which the open string is symmetrically extended.
(For more details see the literature on strings, e.g., \ci{Strings}.)

If we generalize the action (\ref{4A.1}) to $C$-space, we obtain
\be
     I[X] = {\kappa\oo 2} \int \dd \tau \, \dd \sigma \, ({\dot X}^M {\dot X}^N-
  X'^M X'_N)G_{MN}
\lbl{4A.2}
\ee
where $\kappa$ is the generalized string tension.
Taking 4-dimensional spacetime, there are $D=2^4 = 16$ dimensions of the
corresponding $C$-space.
Its signature $(+++...- - - ...)$ has 8 plus and 8 minus signs.
The variables $X^M$ are components of a polyvector $X$ expanded
according to eq. (\ref{2.11a}) and they depend on two parameters
$\tau$ and $\sigma$. From the point of view of $C$-space the variables
$X^M (\tau,\sigma)$ describe an object with two intrinsic
dimensions, that is, a 2-dimensional `world sheet' living in a 16-dimensional
$C$-space. Therefore we will keep on talking about `strings'
(that sweep a world sheet).

Let  us consider the case of an open string satisfying the boundary condition
$X'^M = 0$ at $\sigma = 0$ and $\sigma = \pi$. Then we can make
the expansion
\be
      X^M (\tau,\sigma) = \sum_{n=-\infty}^\infty X_n^M (\tau)
      \, {\rm e}^{i n \sigma}
\lbl{4A.3}
\ee
where from the reality condition $(X^M)^* = X^M$ it follows
\be
     X_n^M = X_{-n}^M
\lbl{4A.4}
\ee
Inserting (\ref{4A.3}) into (\ref{4A.2}), integrating over $\sigma$ and taking
into account (\ref{4A.4}) we obtain the action expressed in terms of
$X_n^M (\tau)$:
\be
   I[X_n^M] = {\kappa'\oo 2} \int \dd \tau \, \sum_{n=-\infty}^\infty
     ({\dot X}_n^M {\dot X}_n^N - n^2 X_n^M X_n^N)G_{MN}
\lbl{4A.5}     
\ee
where $\kappa' = 2 \pi \kappa = 1/\alpha'$. This is just the action of infinite number of
harmonic oscillators.

The Hamiltonian corresponding to the action (\ref{4A.5}) is
\be
    H = {1\oo 2} \sum_{n=-\infty}^{\infty}  \left ( {1\oo \kappa'}
    P_n^M P_{nM} + \kappa' \, n^2 X_n^M X_{nM} \right )
\lbl{4A.6}
\ee
Let us introduce
\bear
     &&a_n^M = {1\oo \sqrt{2}} \left ( {1\oo \sqrt{\kappa'}} P_n^M
     - i n \sqrt{\kappa'} \, X_n^M \right ) \nonumber \\
     &&{a_n^M}^\dagger = {1\oo \sqrt{2}} \left ( {1\oo \sqrt{\kappa'}} P_n^M
     + i n \sqrt{\kappa'} \, X_n \right )
\lbl{4A.7}
\ear
We see that
\be
    a_{-n}^M = {a_n^M}^\dagger
\lbl{4A.7a}
\ee
Rewriting $H$ in terms of $a_n^M$, ${a_n^M}^\dagger$ we obtain
\be
    H = {1\oo 2} \sum_{n=-\infty}^\infty ({a_n^M}^\dagger a_{nM} +
    a_{nM} {a_n^M}^\dagger ) = \sum_{n=1}^{\infty}
    ({a_n^M}^\dagger a_{nM} + a_{nM} {a_n^M}^\dagger )
    + {1\oo {2 \kappa'}} P_0^M P_{0 M}
\lbl{4A.8}
\ee

Upon quantization we have
\be
     [X_n^M,P_{nN}] = i {\delta^M}_N \quad {\rm or} \quad 
     [X_n^M,P_n^N] = i G^{MN}
\lbl{4A.9}
\ee
and
\be
     [a_n^M,a_{nN}^\dagger] = n {\delta^M}_N \quad {\rm or} \quad 
     [a_n^M,{a_n^N}^\dagger] = n G^{MN}
\lbl{4A.10}
\ee

In order to construct the Fock space of excited states, one has first
to define a {\it vacuum state}. There are two possible choices
\ci{PseudoHarm}.

{\it Possibility I}. Conventionally, the vacuum state is defined
according to
\be
      a_n^M |0 \rangle = 0 \; , \quad n \ge 1
\lbl{4A.11}
\ee
and the excited part of the Hamiltonian
$H_{\rm exc} = H-(1/\kappa') P_0^M P_{0M}$, after using (\ref{4A.10}) and 
(\ref{4A.11}) is
\be
    H_{\rm exc} = \sum_{n=-\infty}^\infty ({a_n^M}^\dagger a_{nM} +{n \oo 2}D)
      = 2  \sum_{n=1}^\infty ({a_n^M}^\dagger a_{nM} +{n\oo 2} D) 
\lbl{4A. 12}
\ee
$$D= {\delta^M}_M = G^{MN} G_{MN}$$
Its eigenvalues are all positive\footnote{This is so even for those components
$a_n^M$ that belong to negative signature: negative sign of a term in
${a_n^M}^\dagger a_{nM}$ is compensated by negative sign in the commutation
relation (\ref{4A.10}).} and there is the non vanishing zero point
energy. But there exist negative norm states.

{\it Possibility II}.  Let us split $a_n^M = (a_n^A, a_n^{\bar A})$ where
the indices $A,~{\bar A}$ refer to the components with positive and negative
signature, respectively, and let us define vacuum according to
\be
    a_n^A |0 \rangle = 0 \; , \qquad {(a_n^{\bar A})}^\dagger |0 \rangle = 0
    \; , \quad n \ge 1
\lbl{4A.13}
\ee
Using (\ref{4A.10}) we obtain the Hamiltonian in which the annihilation
operators, defined according to eq.(\ref{4A.13}), are on the right:
\be
    H_{\rm exc} = 2 \sum_{n=1}^\infty ({a_n^A}^\dagger a_{nA} + {n\oo 2} D_{+}
       + a_{n{\bar A}} {a_n^{\bar A}}^\dagger - {n\oo 2} D_{-})
\lbl{4A.14}
\ee
where
\be
    D_{+} = {\delta_A}^A \; , \qquad  D_{-} = {\delta _{\bar A}}^{\bar A}
\lbl{4A.14a}
\ee
are, respectively, the number of positive and negative signature dimensions
and $D=D_{+} + D_{-} = {\delta_M}^M$ the total number of dimensions of Clifford
space.
{\it There are no negative norm states.}

Since in Clifford space the number of positive and negative signature
components is the same, i.e., $D_{+}=D_{-}$, the above Hamiltonian has
vanishing zero point energy:
\be H_{\rm exc} = 2 \sum_{n=1}^\infty ({a_n^A}^\dagger a_{nA} 
       + a_{n{\bar A}} {a_n^{\bar A}}^\dagger)
\lbl{4A.15}
\ee
Its eigenvalues can be positive or negative, depending on which components
(positive or negative signature) are excited.

An immediate objection could arise at this point, namely, that since
the spectrum of the Hamiltonian is not bounded from below, the system
described by $H$ of eq.(\ref{4A.14}) or (\ref{4A.15}) is unstable. This
objection would only hold if the kinetic terms ${\dot X}_n^M {\dot X}_{nM}$
in the action (\ref{4A.5}) (or the terms $P_n^M P_{nM}$ in the Hamiltonian
(\ref{4A.6})) were all positive, so that negative eigenvalues of $H$ would come
from the negative potential terms in $n^2 X_n^M X_{nM}$. But since our
metric is pseudo-Euclidean, whenever a term in the potential is negative,
also the corresponding kinetic term is negative. Therefore, the
acceleration corresponding to negative signature term is proportional to
the {\it plus} gradient of potential (and not to the minus gradient of
potential as it is the case for positive signature term); such system is
{\it stable} if the potential has {\it maximum}, i.e., if it has an upper bound
(and not a lower bound). The overall change
of sign of the action (Lagrangian) has no influence on the equations of
motion (and thus on stability).

In the bosonic string theory based on the ordinary definition of
vacuum (Possibility I) and formulated in $D$-dimensional spacetime
with signature $(+ - - - ... - - -)$ there are negative norm states,
unless $D=26$. Consistency of the string theory requires extra dimensions,
besides the usual four dimensions of spacetime.

My proposal is that, instead of adding extra dimensions to spacetime,
we can start from 4-dimensional spacetime $M_4$ with signature
$(+ - - -)$ and consider the Clifford space ${\cal C}_{M_4}$ ($C$-space)
whose dimension is 16 and signature $(8+,8-)$. {\it The necessary extra
dimensions for consistency of string theory are in $C$-space.} This
also automatically brings {\it spinors} into the game. It is an old
observation that spinors are the elements of left or right ideals of
Clifford algebras \ci{Teitler}--\ci{Lounesto} (see also a very lucid and
systematic recent exposition in refs.\,\ci{Mankoc}). In other words, spinors
are particular sort of
polyvectors \ci{PavsicBook}. Therefore, the string coordinate
polyvectors contain spinors.
This is an alternative way of introducing spinors into the string theory
\ci{PavsicBook,Castro-Pavsic}. An attempt to achieve a deeper understanding of
the structure of supersymmetry has beeb undertaken in refs.\ci{Gates}.
           
Let the constraints (\ref{4A.1a}),(\ref{4A.1b}) be generalized to $C$-space.
So we obtain
\be
    \Pi^M \Pi_M \approx 0 \; , \qquad \Pi^M = P^M + {X'^M\oo 2 \pi \alpha'}
\lbl{4A.16}
\ee
Using (\ref{4A.3}) and expanding the momentum $P^M (\sigma)$ according to
\be
     P^M = \sum_{n=-\infty}^\infty P_n^M {\rm e}^{i n \sigma}
\lbl{4A.17}
\ee
we can calculate the Fourier coefficients of the constraint (called Virasoro
generators):
\be
    L_n = {\pi \alpha'\oo 2} \int_{-\pi}^\pi \dd \sigma \, {\rm e}^{-in \sigma}
    \, \Pi^M \Pi_M = {1\oo 2} \sum_{r=-\infty}^\infty a_{-r}^M a_{r-n}^N G_{MN}
\lbl{4A.18}
\ee

Let us now calculate the commutators  of Virasoro generators. If $m \neq -n$
the commutators can be straightforwardly calculated. The result is
\be
    [L_n, L_m] = {1\oo 2} (n-m) \sum_{r=-\infty}^\infty a_r^M a_{(n+m-r)M}
    = (n-m) L_{n+m} \; , \quad m \neq -n
\lbl{B1}
\ee
For $m= -n$ we have to bear in mind that $a_{-r}^M$ and $a_r^M$ due to
eqs.(\ref{4A.7a}) and (\ref{4A.10}) do not commute. Therefore,
if we put the operators which annihilate the vacuum according
to eq.(\ref{4A.13}) on the right, we obtain the following expression:
\bear
   [L_n, L_{-n}] &=& n \sum_{r=-\infty}^\infty a_r^M a_{-r M} \nonumber \\
   &=& n \left ( a_0^M a_{0 M} + \sum_{r=1}^\infty (a_{-r}^A a_{rA}
   + a_{-r}^{\bar A} a_{r {\bar A}} + a_r^A a_{-rA} +
   a_r^{\bar A} a_{-r {\bar A}} ) \right ) \nonumber \\
   &=& n \left ( a_0^M a_{0 M} + \sum_{r=1}^\infty (2 a_{-r}^A a_{rA}
   + r D_{+} + 2 a_r^{\bar A} a_{-r {\bar A}} - r D_{-} ) \right )
   \nonumber \\
   &=& 2 n L_0 + n  \sum_{r=1}^\infty r (D_{+} - D_{-}) = 2 n L_0 \; , 
   \quad m = -n
\lbl{B2}
\ear
where in the last step we have taken into account that in $C$-space the number
of positive and negative signature dimensions is the same, i.e.,
$D_{+} = D_{-}$.

Combining together eqs. (\ref{B1}) and (\ref{B2}) we find the following
relation   
\be
    [L_m, L_n] = (m-n) L_{n+m}
\lbl{4A.19}
\ee 
which holds for arbitrary positive or negative integers $n$ and $m$.
There are no central terms. The terms which arise after normal
ordering the operators in eq. (\ref{B2}) have opposite sign for positive
and negative signature components,
and thus cancel out. The algebra of Virasoro generators is thus closed,
which automatically assures consistency of quantum string theory formulated
in 16-dimensional Clifford space generated by the spacetime vectors
$\gamma_\mu$.

\section{Conclusion}

We have shown that generalizing physics from 4-dimensional spacetime
to Clifford space
($C$-space) has promising consequences for quantum field theory and
string theory. In order to avoid inconsistencies,
the definition of $C$-space metric as the scalar product of
two basis elements has to involve reversion. The metric so defined
has signature $(+++...---...)$ ( eight times plus, eight times minus).
In QFT formulated in $C$-space the vacuum energy vanishes, if the vacuum state
is defined in a very natural and straightforward way as proposed
by Jackiw \ci{Jackiw} (see also \ci{PseudoHarm}). Therefore, there is
no cosmological constant problem in such a theory. Generalizing the
4-dimensional target space (in which a string lives) to a 16-dimensional
Clifford space, we have found that the (quantum) algebra of Virasoro
generators has no central terms. String theory is consistent if
formulated in 16-dimensional Clifford space. This automatically
brings fermions into the game, since fermions are the elements of
left and right ideals of a Clifford algebra. Therefore the $C$-space
formulation of string theory is an alternative to the usual 
superstring formulation which involves 10 extra dimensions
of target space and Grassmann odd variables. While in the usual string theory
one has to go beyond the 4-dimensional spacetime, and then study how
to compactify the unobservable extra dimensions, in the proposed new theory we
remain in 4-dimensional spacetime. The "extra dimensions" reside
in Clifford space, and they have a physical interpretation as
providing a description of extended object \ci{PavsicArena, PavsicBook}.
We have thus found a very promising outline of QFT and string theory
which deserves further studies.

\end{document}